\newcommand{\cv}[1]{}
\newcommand{\av}[1]{#1}

\cv{
\documentclass{article}
\pdfpagewidth=8.5in
\pdfpageheight=11in
\usepackage{ijcai23}
\newcommand{\citet}[1]{{\citeauthor{#1}~\shortcite{#1}}}
}

\av{\documentclass[letter]{article}
  \usepackage[totalwidth=420pt,totalheight=625pt]{geometry}
  \date{}
  \usepackage{natbib}
  \let\cite=\citep
}

\usepackage{times}
\usepackage{soul}
\usepackage{url}
\usepackage[hidelinks]{hyperref}
\usepackage[utf8]{inputenc}
\usepackage[small]{caption}
\usepackage{graphicx}
\usepackage{amsmath}
\usepackage{amsthm}
\usepackage{booktabs,dcolumn}
\usepackage{algorithm}
\usepackage{algorithmic}
\usepackage[switch]{lineno}

\usepackage{tikz}
\usetikzlibrary{fit}
\usetikzlibrary{arrows.meta,arrows}
\usepackage{tikz-3dplot}
\tikzset{A/.style={draw, circle, inner sep=0pt,
    minimum size=12pt, fill=white}}
\tikzstyle{B}=[draw=black!60] %
\tikzstyle{C}=[draw=black!60]
\tikzset{D/.style={coordinate}}
\tikzset{E/.style={coordinate}}

\tikzset{L/.style={draw=red, circle, inner sep=0pt,
    minimum size=12pt, fill=white}}
\tikzset{R/.style={draw=azure, circle, inner sep=0pt,
    minimum size=12pt, fill=white}}
\tikzset{T/.style={draw=ao, circle, inner sep=0pt,
    minimum size=12pt, fill=white}}
\tikzset{Q/.style={draw=magenta, circle, inner sep=0pt,
    minimum size=12pt, fill=white}}

\tdplotsetmaincoords{120+180+5}{120+15}

\usepackage{xcolor}
\usepackage{url}
\usepackage{xspace}
\usepackage{hyperref}
\usepackage{amsfonts}
\usepackage{cleveref}
\usepackage{enumitem}
\setlist[itemize]{leftmargin=8pt,itemindent=*,noitemsep,topsep=2pt, partopsep=2pt}
\setlist[enumerate]{leftmargin=14pt,itemindent=*,noitemsep,topsep=2pt, partopsep=2pt}

\newcommand{\hy}{\hbox{-}\nobreak\hskip0pt}

\definecolor{ijcaired}{HTML}{D22817}
\definecolor{azure}{rgb}{0.0, 0.5, 1.0}
\definecolor{applegreen}{rgb}{0.55, 0.71, 0.0}
\definecolor{ao}{rgb}{0.0, 0.5, 0.0}
\definecolor{auburn}{rgb}{0.43, 0.21, 0.1}
\definecolor{byzantine}{rgb}{0.74, 0.2, 0.64}
\definecolor{copper}{rgb}{0.72, 0.45, 0.2}
\definecolor{darkbrown}{rgb}{0.4, 0.26, 0.13}
\definecolor{darkchampagne}{rgb}{0.76, 0.7, 0.5}
\definecolor{darkmagenta}{rgb}{0.55, 0.0, 0.55}
\definecolor{darkorange}{rgb}{1.0, 0.55, 0.0}
\definecolor{deepsaffron}{rgb}{1.0, 0.6, 0.2}
\definecolor{lipicsYellow}{rgb}{0.99,0.78,0.07}
\definecolor{fluorescentyellow}{rgb}{0.8, 1.0, 0.0}
\definecolor{goldenyellow}{rgb}{1.0, 0.87, 0.0}
\definecolor{goldenrod}{rgb}{0.85, 0.65, 0.13}
\definecolor{lavenderindigo}{rgb}{0.58, 0.34, 0.92}
\definecolor{lavendermagenta}{rgb}{0.93, 0.51, 0.93}
\definecolor{magenta}{rgb}{1.0, 0.0, 1.0}

\newtheorem{theorem}{Theorem}
\newtheorem{observation}{Observation}

\cv{
\pdfinfo{
	/TemplateVersion (IJCAI.2023.0)
}
}

\newcommand{\SB}{\{\,}%
\newcommand{\SM}{\;{:}\;}%
\newcommand{\SE}{\,\}}%

\newcommand{\GGG}{\mathcal{G}}

\newcommand{\n}[1]{\lnot #1}
\newcommand{\ass}{\alpha}

\usepackage[textwidth=18mm,textsize=tiny]{todonotes}
\definecolor{stefan}{rgb}{.85,.66,0}

\usepackage{wrapfig}

\newcommand{\NP}{\textrm{NP}}   
\newcommand{\coNP}{\textrm{co-NP}}   

\newcommand{\lv}[1]{#1}

\title{Co-Certificate Learning with SAT Modulo Symmetries}

\cv{
 \author{
 Markus Kirchweger%
 \and
 Tomáš Peitl%
 \And
 Stefan Szeider%
 \affiliations
 Algorithms and Complexity Group, TU Wien, Austria\\
 \emails
 \{mk, peitl, sz\}@ac.tuwien.ac.at
}
}

\av{
\author{%
 Markus Kirchweger,
 Tomáš Peitl, and 
 Stefan Szeider\\[4pt]
\small Algorithms and Complexity Group\\[-3pt]
\small TU Wien, Vienna, Austria\\[-3pt] 
\small \texttt{ \{mk|peitl|sz\}@ac.tuwien.ac.at }
}}

\begin{document}

\maketitle
\av{\thispagestyle{empty}}
\begin{abstract}
  We present a new SAT-based method for generating all graphs up to
  isomorphism that satisfy a given co-NP property. Our method extends
  the SAT Modulo Symmetry (SMS) framework %
  with a technique that we call co-certificate learning. If SMS
  generates a candidate graph that violates the given  co-NP property,
  we obtain a certificate for this violation, i.e., `co-certificate'
  for the co-NP property. The co-certificate gives rise to a clause
  that the SAT solver, serving as SMS's backend, learns as part of
  its CDCL procedure. We demonstrate that SMS plus co-certificate
  learning is a powerful method that allows us to improve the
  best-known lower bound on the size of Kochen-Specker vector systems,
  a problem that is central to the foundations of quantum mechanics
  and has been studied for over half a century. Our approach is orders
  of magnitude faster and scales significantly better than a recently proposed
  SAT\hy based method.%

\end{abstract}

\section{Introduction}

SAT modulo symmetries (SMS) is a recently proposed framework that
brings efficient symmetry breaking to conflict-driven (CDCL) SAT
solvers and has achieved state-of-the-art results on several
symmetry-rich combinatorial search problems, enumerating or proving
the non-existence of graphs, planar graphs, directed graphs, and
matroids with particular
properties~\cite{KirchwegerSzeider21,KirchwegerScheucherSzeider22,KirchwegerScheucherSzeider23}. 
 In
this paper, we propose to extend SMS to a class of problems that do
not admit a succinct SAT encoding because they involve quantifier
alternation: where we are asked to find a combinatorial object (the
existential part) that has some $\coNP$-complete property (the
universal part, stated as `all candidate polynomial-size witnesses
fail').  We call such problems \emph{alternating search} problems; a
simple concrete example of an alternating search problem is the
well-studied question posed by~\citet{Erdos67}, of finding a smallest
triangle-free graph that is not properly $k$-colorable, for a fixed
$k\geq 3$.  Encoding the non\hy $k$\hy colorability property for
$k\geq 3$ into a family of polynomially sized propositional formulas
is impossible unless $\NP=\coNP$, since checking $k$\hy colorability
is $\NP$\hy complete \cite{Karp72}.

SMS has some advantages over alternative methods such as isomorphism-free exhaustive enumeration by canonical construction path~\cite{McKay98} as implemented in tools like Nauty~\cite{McKayPiperno14}, or different symmetry-breaking methods for SAT.
The former can very efficiently generate all objects of a given order, but is very difficult to integrate with complex constraints and learning, while the latter is either intractable (full `static' symmetry breaking~\cite{CodishGIS16,ItzhakovC15}, which requires constraints of exponential size), or ineffective (partial static symmetry breaking~\cite{CodishMillerProsserStuckey19}).
Because SMS strikes a better balance than these other methods, with both native constraint reasoning and learning as well as effective and efficient symmetry breaking, we chose it as our basis for alternating search.

We call our new method SMS plus \emph{co-certificate learning (CCL)},
and it works as follows.  We run an SMS solver on an encoding of the
existential part of our alternating search problem, giving us a model
that corresponds to a \emph{solution candidate}.  In the context of
our running example, this candidate would be a triangle-free graph.
Next, we test the \coNP{} property (non-$k$-colorability).  If the graph
is not colorable, we have found a solution (and we may or may not
proceed to enumerate all solutions as with any other SAT encoding).
If it is
colorable, we find a coloring, which is a \emph{co-certificate} of the
graph \emph{not} having our desired non-colorability property.  The
co-certificate gives rise to a clause that is learned by the SMS
solver and prevents any solutions with the same co-certificate
(colorable by the same coloring), and search resumes.  We describe the
method in full detail in Section~\ref{section:cocert}---for now
suffice it to say that we essentially apply SMS incrementally with
respect to the learned co-certificates.

Co-certificate learning is a general method that applies to any alternating search problem, but it is much easier to explain on a concrete example like graph coloring.
In Sections~\ref{sec:KS} and \ref{sec:ks-new-bounds}, we present a more involved application of CCL, to the existence of \emph{Kochen-Specker (KS) systems}, a combinatorial object that features in the proof of the Bell-Kochen-Specker theorem in quantum mechanics.
With SMS+CCL we achieve an orders-of-magnitude speed-up over a different recently proposed SAT-based approach to the KS problem, and we prove that any KS system must have at least $24$ vectors.

\section{Preliminaries}

For a positive integer~$n$, we write $[n] = \{1,2,\dots,n\}$.
We assume familiarity with fundamental notions of propositional logic~\cite{Prestwich09}. %
In this paper we are presenting a general method for alternating combinatorial search problems on particular examples with graphs; below we review basic notions from graph theory relevant to our discussion.

\paragraph{Graphs.}

All considered graphs are undirected and simple (i.e.,
without parallel edges or self-loops). A \emph{graph} $G$ consists of
set $V(G)$ of vertices and a set $E(G)$ of edges; we denote the edge
between vertices $u,v\in V(G)$ by $uv$ or equivalently $vu$. The \emph{order} of a graph $G$ is the number of its vertices,~$|V(G)|$. We write
$\GGG_n$ to denote the class of all graphs with $V(G) = [n]$.  The \emph{adjacency matrix} of a
graph $G \in \GGG_n$, denoted by $A_G$, is the $n\times n$ matrix where the element at row $v$ and column $u$, denoted by $A_G[v][u]$, is $1$ if $vu \in E$ and
$0$ otherwise. %

\paragraph{Isomorphisms.}
For a permutation $\pi : [n] \rightarrow [n]$, $\pi(G)$ denotes the graph obtained from $G\in \GGG_n$ by the
permutation $\pi$, where $V(\pi(G)) = V(G) = [n]$ and
$E(\pi(G))=\SB \pi(u)\pi(v)\SM uv \in E(G) \SE$.
Two graphs $G_1,G_2\in \GGG_n$ are \emph{isomorphic} if there is a
permutation $\pi : [n] \rightarrow [n]$ such that $\pi(G_1)=G_2$; in this case $G_2$ is an \emph{isomorphic copy} of $G_1$.

\paragraph{Coloring.}
A \emph{(proper) $k$-coloring} of a graph $G$ is a map $c : V(G) \rightarrow [k]$ with the property that if $uv \in E(G)$, then $c(u) \neq c(v)$ (adjacent vertices have different colors).

\paragraph{Partial graphs.}
A \emph{partially defined graph} is a graph $G$ where
$E(G)$ is split into two disjoint sets~$D(G)$ and~$U(G)$.  $D(G)$
contains the \emph{defined} edges, $U(G)$ contains the \emph{undefined} edges.  A
(\emph{fully defined}) graph is a partially defined graph $G$ with
$U(G)=\emptyset$.
A partially defined graph $G$ can be \emph{extended} to a
graph $H$ if  $D(G) \subseteq E(H) \subseteq D(G) \cup
U(G)$.

\paragraph{SAT Modulo Symmetries (SMS).}

SMS is a framework that augments a CDCL (conflict-driven clause learning) SAT solver~\cite{FichteHLS23,MarquessilvaLynceMalik09} with a custom propagator that can reason about symmetries, allowing to search modulo isomorphisms for graphs in $\GGG_n$ which satisfy constraints described by a propositional formula.
During search the SMS propagator can trigger additional conflicts on
top of ordinary CDCL and consequently learn \emph{symmetry-breaking clauses}, which exclude isomorphic copies of graphs. More precisely,
only those copies which are lexicographically minimal (\emph{canonical}) by
concatenating the rows of the adjacency matrix are kept. A key
component is a minimality check, which decides whether a partially
defined graph can be extended to a minimal graph; if it cannot, a
corresponding clause is learned.  For a full description of SMS, we
refer to the original work where the framework was introduced~\cite{KirchwegerSzeider21}.

\section{Co-Certificate Learning (CCL)}\label{section:cocert}

In this paper we propose to extend SMS to alternating search problems.
The way we do this is by applying SMS incrementally.  In this section we shall explain the procedure by showing how it would play out in our running example of finding a triangle-free graph with $n$ vertices and without a $k$-coloring.
Throughout this section we will talk about propositional formulas that encode graphs with the vertices $[n]$ using the variables $e_{u,v}$, where $u,v \in [n]$.
Any truth assignment $\ass$ defines the (possibly partially defined)
graph $G_\ass$ with  $V(G_\ass)= [n]$ and $E(G_\ass)=\{uv \;:\; \ass(e_{u,v}) = 1\})$.

\av{\sloppypar}
First, we generate a triangle-free graph, which is obtained from a SAT
solver as a model of a (polynomial-size) propositional formula
encoding triangle-freeness---the clauses \[ \n{e_{u,v}} \vee \n{e_{v,w}} \vee \n{e_{u,w}} \text{ for } u,v,w \in V, u<v<w \] forbid all triangles. 
We feed the candidate graph into a custom
coloring algorithm and test whether it is $k$-colorable.  We will only
deal with small enough graphs that the time cost of coloring, though
exponential in the worst case, is going to be negligible.
On the other hand, it could be the case that candidate graphs have many colorings, and the \emph{choice} of a particular coloring might matter much more. We will elaborate on this in Section~\ref{sec:KS-encoding}. %

Returning to our candidate graph, if it is not
colorable, we obtain a solution. Depending on whether we want just one
or all solutions we can stop or add a clause blocking this particular
solution and resume.  If, on the other hand, the graph is
$k$-colorable, we obtain a certificate for this in the form of a
coloring $c : [n] \rightarrow [k]$. %
We learn and add a
\emph{coloring clause} \[C_c = \bigvee_{ u < v \in [n], c(u) = c(v)} e_{u,v}. \]  This clause says that future generated graphs
must not be colorable by the coloring $c$.

Finally, we go back to step one and generate another graph, this time respecting the newly learned coloring clause.
We repeat this until either a non-$k$-colorable graph is found or the formula becomes unsatisfiable.

\begin{figure}
\centering
		\begin{tikzpicture}
			\node[draw,rounded corners=2] (CDCL) at (0, 0) {CDCL Solver};
			\node[draw,rounded corners=2] (fullobj) at (2, 1) { Solution Candidate};
			\node[draw,rounded corners=2] (mincheck) at (0, 2) {Symmetry Check};

			\node[draw=ijcaired,thick,rounded corners=2] (conp) at (5, 0) {co-NP Check};

			\node[draw=azure,thick,rounded corners=2] (sol) at (5, 2) {Solution};

			\node[draw,dashed,inner sep=2mm,label=below:SMS,fit=(mincheck) (CDCL) (fullobj) (CDCL)] (graphgen) {};

			\draw (CDCL) edge[-latex,bend left] (mincheck);
			\draw (mincheck) edge[-latex,bend left] (CDCL);

			\draw (CDCL) edge[-latex,bend left=5] (fullobj);
			\draw (fullobj) edge[-latex,bend left=20] (conp);
			\draw (conp) edge[-latex] node[above] {\small co-certificate} node[below] {\small blocking clause} (CDCL);

			\draw (conp) edge[-latex] (sol);
		\end{tikzpicture}
	\caption{
		Co-certificate learning = SMS + \coNP{} property check.
	}
	\label{fig:CCL}
\end{figure}
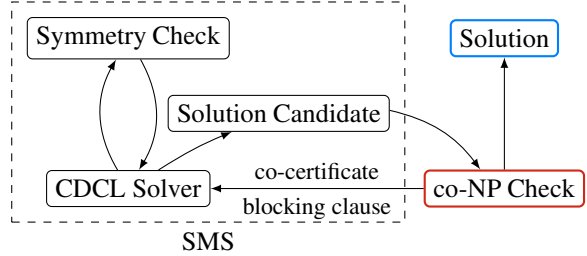

The full process is illustrated in Figure~\ref{fig:CCL}.
The crucial innovation is the
learning of coloring clauses.  Thanks to the fact that a single
coloring is valid for multiple graphs, we do not have to visit every
candidate graph in the search as we would with a naive method that does not interleave graph generation with coloring.
Instead, it is sufficient to `visit enough colorings,' in the sense that
once a set of colorings (and their corresponding clauses) that jointly
color all colorable graphs of order $n$ is generated (and any
non-colorable graphs are found), the formula immediately becomes
unsatisfiable.

We call this process \emph{co-certificate learning (CCL)}, after the fact that each learned coloring is a `co-certificate' that the graph does \emph{not} have the desired non-colorability property.
It follows immediately that such a co-certificate learning can be formulated
for any combinatorial search problem where the sought-for object has a \coNP{} property.
Here we chose to present CCL on the conceptually simpler triangle-free non-colorable problem,
but in Section~\ref{sec:ks-new-bounds} we will present a more involved showcase
for SMS+CCL, one where it goes significantly beyond the state of the art.

We have glossed over this until now, but CCL is not running in an ordinary SAT solver but on top of SMS.
This means every now and then we hook into the SAT solver and check whether the currently considered graph, which may not be fully defined, is canonical.
If it is not canonical, we learn a clause that causes the solver to immediately backtrack.
This is the same as ordinary SMS, and likewise, all candidate graphs produced are canonical and thus unique w.r.t.\ isomorphism.
As a consequence, the set of co-certificates (colorings) that we need to learn is in fact only required to work for canonical graphs, and not all graphs of order $n$.

\subsection{Related Methods}

In the last 20 years many different methods have been developed to extend SAT solvers with external algorithms in order either to improve performance or to grant SAT solvers the ability to solve non-Boolean problems.
While these methods are superficially similar, the specifics of how the collaboration is carried out and the type of external algorithm differ significantly.
Among the best known are \emph{SAT modulo Theories (SMT)}~\cite{SMT} and \emph{lazy clause generation (LCG)}~\cite{LCG}; other methods that fall into this category are the \emph{SAT+CAS} system~\cite{ZulkoskiBHKCG17}, where the SAT solver communicates with a computer algebra system (CAS), and \emph{counterexample-guided abstraction-refinement (CEGAR)} solvers for quantified Boolean formulas (QBF)~\cite{JanotaKlieberMarquessilvaClarke12}, which typically use two or more communicating SAT solvers.

Our approach, SMS+CCL, also belongs to this family.
SMS delegates the symmetry breaking to a specialized algorithm; CCL delegates the co-certificate search to another specialized algorithm.
The most similar approach to CCL is an abstraction-based algorithm for
2QBF due to~\citet{JanotaMarquessilva11}, which solves formulas of the
form $\exists X \forall Y (\lnot \phi)$ where $\phi$ is a CNF
formula. The part in common with our method is that solution candidates are computed and excluded by adding additional clauses if necessary.
What makes SMS+CCL stand out is the high degree of integration of its
components and the fact that the existential and universal parts are handled separately.
Both symmetry breaking and co-certificate learning are carried out by our code, tightly integrated into a single binary, having access to the entire state of the solver.
Another hallmark of our approach is that we focus on graphs (and, more
generally, combinatorial objects), so our symmetry-breaking and
co-certificate learning do not operate on a propositional encoding (as
one is forced to when solving pure QBFs, for example). However, they understand the high-level structure and can thus be more specialized and effective.
At the same time, our method remains general enough to apply to new alternating search problems with only a little implementation effort.

\subsection{Comparison with Alternatives}

For the remainder of this section, we will present an experimental comparison of SMS+CCL to various alternative methods on the task of finding non-$3$-colorable triangle-free graphs with 10--14 vertices, showing how we can visit fewer candidate graphs and learn small sets of colorings thanks to SMS and CCL.
The methods we compare are:
\begin{itemize}%
	\item \av{\sloppypar}Nauty~\cite{McKayPiperno14}: enumeration modulo isomorphism of triangle-free graphs via Nauty's built-in \texttt{geng -t}, followed by filtering out non-$3$-colorable graphs by our custom code;
	\item PSS+CCL: enumeration of triangle-free graphs with a SAT solver and with \emph{partial static symmetry breaking}~\cite{CodishMillerProsserStuckey19}, interleaved with CCL for 3-colorability (same custom code as above, custom implementation of the symmetry breaking constraints);
	\item SMS+CCL: as described above (using the same SAT solver as PSS).
\end{itemize}

Another alternative that we did not consider for this comparison is to use full static symmetry breaking, which produces a constraint of exponential size and is known to scale to no more than $12$ vertices~\cite{CodishGIS16}. %
 
\begin{table}
	\centering
	\begin{tabular}{@{}rrrr@{}}
		\toprule
		$n$ & Nauty        & PSS+CCL       & SMS+CCL \\
		\midrule                                                       
		10  & 3.06         & 0.02          & 0.08    \\
		11  & 29.23        & 0.10          & 0.15    \\
		12  & 375.38       & 1.67          & 0.42    \\
		13  & 6507.02      & 774.39        & 2.95    \\
	    14  & $\sim$ 2 days& $>$4 days        & 147.57  \\
		 \bottomrule
	\end{tabular}
	\caption{
		Running times in seconds of different approaches to the $\Delta$-free non-colorable problem.
		The time for Nauty includes postprocessing to filter out 3-colorable graphs, using the same code as for the other methods.
	}
	\label{table:CCL-vs-others-time}
\end{table}

From Table~\ref{table:CCL-vs-others-time} we can see that CCL provides a speed-up of many orders of magnitude over isomorphism-free exhaustive generation with Nauty followed by $3$-colorability filtering.
Table~\ref{table:CCL-vs-others-size} explains this speed-up clearly in terms of the number of graphs that need to be enumerated.
Nauty always needs to enumerate all (non-isomorphic) graphs, the number in the first column.
This is many orders of magnitude larger than the number of (fully defined) graphs visited by SMS + CCL: those in the third column.
Even more striking is the tiny number of colorings that are sufficient to color all canonical graphs, shown in the parentheses in the last column of Table~\ref{table:CCL-vs-others-size}.
In the case of $n=14$, a single $3$-coloring covers almost 50000 canonical graphs on average!
Incomplete static symmetry breaking on the other hand, while allowing the use of CCL, is not as good at reducing the search space, as can be witnessed by the running times and the number of graphs enumerated, as well as the total number of graphs that pass the symmetry breaking constraint, shown in Table~\ref{table:CCL-vs-others-size}.

Note that in this comparison we did not tweak the details of the
methods so as to extract every last bit of performance (see, e.g., the
work by \citet{Goedgebeur20} for a high-tech Nauty-based solution to the triangle-free non-colorable problem).
The point of this comparison is to demonstrate the inherent limits of each of these approaches, as well as to compare the `off-the-shelf' versions of every method.

We thus conclude that, at least in the context of alternating problems like graph coloring, co-certificate learning is absolutely crucial to prevent exhaustive enumeration, and SMS
is the symmetry-breaking method that scales best.

Having introduced our general framework of co-certificate learning, we
shall now turn our attention to the interesting multi-faceted
application that is the search for a smallest Kochen-Specker vector system.

\begin{table}
	\centering
	\begin{tabular}{@{}rrrrr@{}}
		\toprule
		$n$ & Nauty      & PSS     & PSS+CCL          & SMS+CCL        \\
		\midrule                                                                          
		10  & $12172$    &  309987 & $171$            & $54(54)$       \\
		11  & $105071$   & 9969561 & $618$            & $147(146)$     \\
		12  & $1262180$  &         & $3668$           & $505(481)$     \\
		13  & $20797002$ &         & $171780$         & $3124(2014)$   \\
		14  & $467871369$&         & $>2\times10^7$   & $85668(9407)$  \\
		 \bottomrule
	\end{tabular}
	\caption{
		1st column: the number of non-isomorphic $\Delta$-free graphs of order $n$; Nauty has to process all of them.
		2nd column: the number of $\Delta$-free graphs with $n$ vertices that pass the PSS constraint of \protect\citet{CodishMillerProsserStuckey19} (we did not compute the missing entries, because $n=11$ already took more than 8 hours).
		This gives an idea of the general effectiveness of PSS vis-à-vis the true number of non-isomorphic graphs.
		3rd and 4th column: the number of graphs visited by PSS+CCL and SMS+CCL respectively, in parentheses of the 4th column the number of colorings (co-certificates) learned by SMS+CCL; contrast this number with the vastly larger total number of graphs in the first column.
	}
	\label{table:CCL-vs-others-size}
\end{table}

\section{Kochen-Specker Systems} \label{sec:KS}

\emph{Kochen-Specker (KS)} vector systems are specific sets of vectors
in at least 3-dimensional space that form the basis of the
Bell-Kochen-Specker Theorem, a central result in the foundations of
quantum mechanics. The existence of a KS vector system shows that
quantum mechanics is in conflict with classical models in which the
result of a measurement does not depend on which other compatible
measurements are jointly performed, a phenomenon known as
\emph{contextuality}~\cite{BudroniEtal22}. In their original 1967
paper, \citet{KochenSpecker67} proposed a 3-dimensional KS vector
system of size 117.  Since then, researchers have been striving to
find smaller KS vector systems and establish lower bounds on their
size.  KS vector systems of a higher dimension $n \geq 3$ are also
considered in the literature, with the additional property that any
pair of vectors belongs to a set of $n$ mutually orthogonal
vectors~\cite{PavicicEtal05}. Higher dimensions allow for smaller KS
systems, whereas the additional property increases the size of
smallest KS systems. In the following we focus on KS vector systems of
dimension~3.

The smallest KS vector system (of dimension 3) known to date is due to
Conway and Kochen and has 31 vectors~\cite{Peres91}. The first lower
bound, of 18, was given by \citet{ArendsOuaknineWampler11}; later it was
improved to 22 by \citet{UijlenWesterbaan16}, and recently to 23 by
\citet{LiBrightGanesh22}.  All these lower bounds were obtained by
computer search methods for undirected graphs associated with KS
vector systems. Since the search space is enormous and further
increases by several orders of magnitude with each increment, the
methods must become more and more sophisticated.
\citeauthor{ArendsOuaknineWampler11} obtain their lower bound with
backtracking search that rules out all KS vector systems with 17 or fewer
vectors; \citeauthor{UijlenWesterbaan16} improved this method by
deriving additional necessary restrictions on the graphs, ruling out
KS vector systems with up to 21 vectors. \citeauthor{LiBrightGanesh22}~propose 
a SAT-based method where the SAT solver interacts with a
computer algebra system (CAS) and prove non-existence of KS systems with 22 vectors.
We will discuss the difference between their and our approach below.
 
We give a lower bound of 24 using SMS+CCL to exclude KS
systems with up to 23 vectors.

\av{\sloppypar}
Next, we define the properties of graphs associated with KS vector systems,
closely following \citeauthor{ArendsOuaknineWampler11}'s approach,
which has also been adopted by subsequent authors
\cite{UijlenWesterbaan16,LiBrightGanesh22}.

A \emph{KS graph}  is a simple undirected graph which is not
010-colorable but embeddable, where these two properties
are defined as follows.
A graph $G$ is \emph{010-colorable} if we can assign  0 or 1 to its
vertices in such a way that (i) no two adjacent vertices are both
assigned 0, and (ii) vertices forming a triangle are not all assigned~1.
$G$ is \emph{embeddable} if its vertices can be mapped to three-dimensional real vectors %
such that adjacent
vertices are orthogonal and there is no collinear pair.
\lv{Figure~\ref{fig:ks31} shows Conway and Kochen's 31-vertex KS graph and
an embedding. 

\begin{figure}[tbh]
  \centering

  \scalebox{0.8}{
  	\begin{tikzpicture}[scale=1.5]
  	\small
  	
  	\draw (0,0) node[coordinate] (origin) {}
  	+(1*360/31: 2cm) node  [style=R] (1) {1}
  	+(2*360/31: 2cm) node  [style=T] (14) {14}
  	+(3*360/31: 2cm) node  [style=R] (3) {3}
  	+(4*360/31: 2cm) node  [style=T] (13) {13}
  	+(5*360/31: 2cm) node  [style=R] (11) {11}
  	+(6*360/31: 2cm) node  [style=Q] (31) {31}
  	+(7*360/31: 2cm) node  [style=T] (19) {19}
  	+(8*360/31: 2cm) node  [style=L] (23) {23}
  	+(9*360/31: 2cm) node  [style=R] (7) {7}
  	+(10*360/31: 2cm) node [style=T] (17) {17}
  	+(11*360/31: 2cm) node [style=L] (22) {22}
  	+(12*360/31: 2cm) node [style=R] (5) {5}
  	+(13*360/31: 2cm) node [style=L] (30) {30}
  	+(14*360/31: 2cm) node [style=Q] (28) {28}
  	+(15*360/31: 2cm) node [style=R] (10) {10}
  	+(16*360/31: 2cm) node [style=Q] (29) {29}
  	+(17*360/31: 2cm) node [style=T] (27) {27}
  	+(18*360/31: 2cm) node [style=R] (6) {6}
  	+(19*360/31: 2cm) node [style=L] (25) {25}
  	+(20*360/31: 2cm) node [style=R] (9) {9}
  	+(21*360/31: 2cm) node [style=T] (18) {18}
  	+(22*360/31: 2cm) node [style=R] (4) {4}
  	+(23*360/31: 2cm) node [style=L] (21) {21}
  	+(24*360/31: 2cm) node [style=R] (2) {2}
  	+(25*360/31: 2cm) node [style=L] (20) {20}
  	+(26*360/31: 2cm) node [style=T] (15) {15}
  	+(27*360/31: 2cm) node [style=L] (24) {24}
  	+(28*360/31: 2cm) node [style=R] (8) {8}
  	+(29*360/31: 2cm) node [style=T] (16) {16}
  	+(30*360/31: 2cm) node [style=L] (26) {26}
  	+(31*360/31: 2cm) node [style=Q] (12) {12}
  	;

  	\draw [style=B] (1) to (12);
  	\draw [style=C] (1) to (14);
  	\draw [style=C] (1) to (21);
  	\draw [style=B] (1) to (30);
  	\draw [style=B,bend left=5] (2) to (16);
  	\draw [style=C] (2) to (20);
  	\draw [style=B] (2) to (21);
  	\draw [style=C] (3) to (13);
  	\draw [style=B] (3) to (14);
  	\draw [style=B,bend left=5] (3) to (23);
  	\draw [style=C] (4) to (18);
  	\draw [style=C] (4) to (21);
  	\draw [style=B,bend right=5] (4) to (27);
  	\draw [style=B,bend right=5] (4) to (29);
  	\draw [style=B] (5) to (15);
  	\draw [style=B] (5) to (16);
  	\draw [style=B] (5) to (22);
  	\draw [style=B, bend right=8] (5) to (23);
  	\draw [style=B, bend left=8] (5) to (27);
  	\draw [style=B] (5) to (30);
  	\draw [style=C] (6) to (14);
  	\draw [style=C] (6) to (25);
  	\draw [style=B] (6) to (27);
  	\draw [style=B] (6) to (31);
  	\draw [style=C] (7) to (17);
  	\draw [style=B] (7) to (18);
  	\draw [style=B] (7) to (23);
  	\draw [style=B] (8) to (16);
  	\draw [style=C] (8) to (24);
  	\draw [style=B] (8) to (25);
  	\draw [style=C] (9) to (18);
  	\draw [style=C] (9) to (25);
  	\draw [style=B] (9) to (28);
  	\draw [style=B] (9) to (30);
  	\draw [style=B] (10) to (19);
  	\draw [style=B] (10) to (23);
  	\draw [style=B] (10) to (24);
  	\draw [style=C] (10) to (28);
  	\draw [style=B] (10) to (29);
  	\draw [style=B] (11) to (13);
  	\draw [style=B] (11) to (16);
  	\draw [style=B] (11) to (26);
  	\draw [style=B] (11) to (28);
  	\draw [style=C] (11) to (31);
  	\draw [style=B] (12) to (17);
  	\draw [style=C] (12) to (19);
  	\draw [style=B] (12) to (26);
  	\draw [style=B] (12) to (30);
  	\draw [style=C] (13) to (22);
  	\draw [style=B] (13) to (28);
  	\draw [style=B,bend left=5] (14) to (23);
  	\draw [style=C] (15) to (20);
  	\draw [style=B] (15) to (22);
  	\draw [style=C] (15) to (24);
  	\draw [style=B,bend right=5] (16) to (21);
  	\draw [style=B] (16) to (23);
  	\draw [style=B] (16) to (25);
  	\draw [style=B] (16) to (26);
  	\draw [style=C] (17) to (22);
  	\draw [style=B] (17) to (26);
  	\draw [style=B] (18) to (23);
  	\draw [style=B] (19) to (20);
  	\draw [style=B] (19) to (23);
  	\draw [style=B] (19) to (31);
  	\draw [style=B] (20) to (31);
  	\draw [style=B] (24) to (29);
  	\draw [style=C] (26) to (29);
  	\draw [style=B] (27) to (29);
  	\draw [style=B,bend right=8] (27) to (30);
  	\draw [style=B] (27) to (31);
  	\draw [style=B] (28) to (30);

  	\end{tikzpicture}
  }
  \vspace{5mm}
  
  \scalebox{0.8}{
  	\begin{tikzpicture}[scale=0.85,  tdplot_main_coords]
  	\small
  	
  	\draw
  	node[D, label={[above=-1mm] $(-2,-2,-2)$}] at (-2,-2,-2) (a) {a} 
  	node[D] (b) {b} 
  	node[D, label={[right] $(-2,2,-2)$}] at (-2,2,-2) (c) {c} 
  	node[D, label={[right] $(-2,2,2)$}] at (-2,2,2) (d) {d} 
  	node[D, label={[left=2mm] $(2,-2,-2)$} ] at (2,-2,-2) (e) {e}
  	node[D, label={[left=2mm] $(2,-2,2)$}] at (2,-2,2) (f) {f} 
  	node[D] (g) {g} 
  	node[D, label={[below=2mm] $(2,2,2)$}] at (2,2,2) (h) {h} 
  	(f)--(h)--(d)
  	(e)--(g)--(c)--(a)--(e)
  	(f)--(e) (h)--(g) (d)--(c)  (f)--(e)
  	;
  	
  	\draw
  	node[E] at (-1,2,2) (dh1) {}
  	node[E] at (0,2,2) (dh2) {}
  	node[E] at (1,2,2) (dh3) {}
  	node[E] at (-1,2,-2) (gc1) {}
  	node[E] at (0,2,-2) (gc2) {}
  	node[E] at (1,2,-2) (gc3) {}
  	node[E] at (-1,-2,-2) (ea1) {}
  	node[E] at (0,-2,-2) (ea2) {}
  	node[E] at (1,-2,-2) (ea3) {}
  	node[E] at (2,-2,-1) (fe1) {}
  	node[E] at (2,-2,0) (fe2) {}
  	node[E] at (2,-2,1) (fe3) {}
  	node[E] at (2,2,-1) (hg1) {}
  	node[E] at (2,2,0) (hg2) {}
  	node[E] at (2,2,1) (hg3) {}
  	node[E] at (-2,2,-1) (dc1) {}
  	node[E] at (-2,2,0) (dc2) {}
  	node[E] at (-2,2,1) (dc3) {}
  	node[E] at (2,-1,-2) (eg1) {}
  	node[E] at (2,0,-2) (eg2) {}
  	node[E] at (2,1,-2) (eg3) {}
  	node[E] at (2,-1,2) (fh1) {}
  	node[E] at (2,0,2) (fh2) {}
  	node[E] at (2,1,2) (fh3) {}
  	node[E] at (-2,-1,-2) (ac1) {}
  	node[E] at (-2,0,-2) (ac2) {}
  	node[E] at (-2,1,-2) (ac3) {}
  	;
  	
  	\draw[help lines]
  	(dh1)--(gc1)
  	(dh2)--(gc2)
  	(dh3)--(gc3)
  	
  	(ea1)--(gc1)
  	(ea2)--(gc2)
  	(ea3)--(gc3)
  	
  	(fe1)--(hg1)
  	(fe2)--(hg2)
  	(fe3)--(hg3)
  	
  	(dc1)--(hg1)
  	(dc2)--(hg2)
  	(dc3)--(hg3)
  	
  	(eg1)--(fh1)
  	(eg2)--(fh2)
  	(eg3)--(fh3)
  	
  	(eg1)--(ac1)
  	(eg2)--(ac2)
  	(eg3)--(ac3)
  	;
  	
  	\draw
  	node[R] at (-1,2,1) (1) {1}
  	node[R] at (-1,2,0) (2) {2} 
  	node[R] at  (0, 2, 1) (3) {3} 
  	node[R] at (-1,2,-1) (4) {4}
  	node[R] at (0, 2, 0)  (5) {5}
  	node[R] at (1, 2, 1) (6) {6}
  	node[R] at (0, 2, -1) (7) {7}
  	node[R] at (1, 2, 0) (8) {8}
  	node[R] at (1,2,-1) (9) {9}
  	node[R] at  (0,2,-2) (10) {10}
  	node[R] at  (2, 2, 0) (11) {11}
  	node[Q] at  (2,2,-2) (12) {12}
  	node[T] at  (-1, 1, -2) (13) {13}
  	node[T] at  (0, 1, -2) (14) {14}
  	node[T] at  (-1, 0, -2) (15) {15}
  	node[T] at  (0, 0, -2) (16) {16}
  	node[T] at  (-1,-1,-2) (17) {17}
  	node[T] at  (0,-1,-2) (18) {18}
  	node[T] at  (0, -2, -2) (19) {19}
  	node[L] at  (2,1,-1)  (20) {20}
  	node[L] at  (2, 1, 0) (21) {21}
  	node[L] at  (2, 0, -1) (22) {22}
  	node[L] at  (2, 0, 0) (23) {23}
  	node[L] at  (2, -1, -1) (24) {24}
  	node[L] at  (2,-1,0) (25) {25}
  	node[L] at  (2,-2,0) (26) {26}
  	node[T] at  (2, 0, -2) (27) {27}
  	node[Q] at  (2,-2,-2) (28) {28}
  	node[Q] at  (2, 2, 2) (29) {29}
  	node[L] at  (2, 0, 2) (30) {30}
  	node[Q] at  (2, -2, 2) (31) {31};

  	\end{tikzpicture}
  }
  
  \caption{ Top: smallest known KS graph with 31 vertices and 71
	  edges, drawn properly 4-colored. 
	Bottom: an embedding of this graph, where each vertex
    corresponds to a vector starting at the center of the
    cube. For instance, the vertices 1 and 30 are adjacent, and hence
    their corresponding vectors $(-1,2,1)$ and $(2, 0, 2)$,
    respectively, are indeed orthogonal (i.e., perpendicular) since
    $-1\cdot 2+ 2 \cdot 0 + 2 \cdot 1=0$.
} \label{fig:ks31} 
\end{figure}
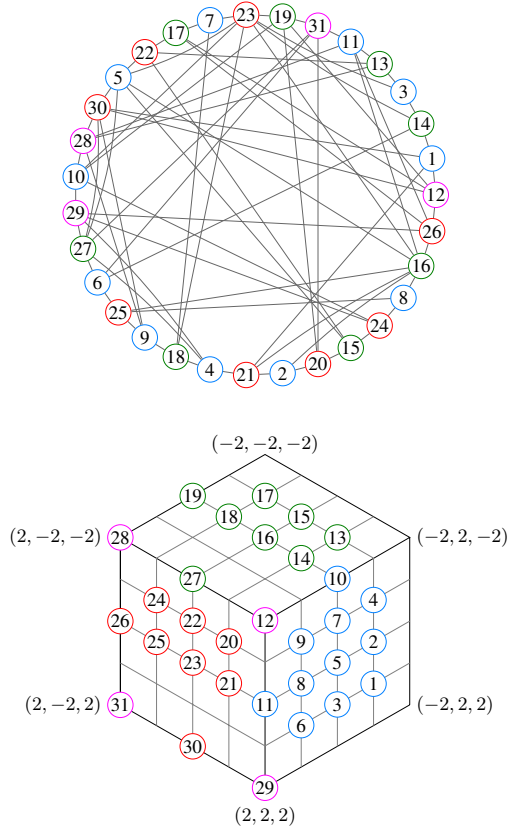}

As shown by \citet{ArendsOuaknineWampler11}, there exists a KS vector system
with $n$ vectors if an only if there exists a KS graph with $n$
vertices.  Moreover, \citeauthor{ArendsOuaknineWampler11}~give the
following necessary properties of any smallest KS graph $G$:

\begin{enumerate}
\item $G$ is square-free (i.e., has no cycle of length 4);
\item $G$ is $4$-colorable;
\item $G$ has minimum degree at least 3;
\item each vertex of $G$ belongs to a triangle. 
\end{enumerate}

Non-010-colorable graphs satisfying the previous four necessary
properties are called \emph{KS candidates}.  They are only
`candidates', because they are not guaranteed to be
embeddable.  All known lower bounds on the size of a KS system were
obtained by enumerating all KS candidates (modulo isomorphisms),
and checking that none are embeddable.  We now turn to the
description of our own process for proving KS lower bounds, which also
follows this two-phase pattern, but differs in our use of the new
SMS+CCL method for the enumeration phase, and in some aspects of the
embeddability check. We note that checking non-010-colorability is
$\coNP$\hy complete \cite{ArendsOuaknineWampler11}, hence this
property is well-suited for our method.

\section{New Lower Bound for KS Systems}
\label{sec:ks-new-bounds}

This section is devoted to a description of our SAT encoding for the search of KS graphs, followed by a discussion of experimental results.
We first produce a formula $F_n$, whose non-010-colorable models correspond to KS candidates; we then enumerate the non-010-colorable models of $F_n$ with SMS+CCL; and finally we check the obtained graphs for embeddability.

\subsection{Encoding and CCL}
\label{sec:KS-encoding}

We use the variables $e_{v,u}$ to indicate whether a certain edge is present (\emph{edge variables}) and the auxiliary variables $t_{v_1,v_2,v_3}$ indicating whether $v_1, v_2, v_3$ forms a triangle. A graph can be extracted from a model of $F_n$ by looking at the assignment of the edge variables. 
Our encoding is almost identical to previous work~\cite{LiBrightGanesh22} with the exception that coloring clauses are learned lazily.
The necessary conditions mentioned at the end of Section~\ref{sec:KS} are encoded as follows:
\begin{enumerate}
	\item $ \lnot e_{v_1,v_2} \lor \lnot e_{v_2,v_3} \lor \lnot e_{v_3,v_4} \lor \lnot e_{v_1,v_4}$ for $v_1, v_2, v_3, v_4 \in [n]$ where $v_i \not = v_j$ if $i \not = j$ and $v_1 < v_2 < v_4$, $v_1 < v_3$, to forbid only one orientation of each 4-cycle.
	\item To express that $G$ must be $4$-colorable, we need additional variables $c_{v,i}$, $v \in [n], i \in [4]$, for the color of each vertex. To ensure each vertex is colored we add the clauses $\bigvee_{i \in [4]} c_{v,i}$ for $v \in [n]$, and to avoid monochromatic edges we add $\lnot e_{v,u} \lor \lnot c_{v,i} \lor \lnot c_{u,i}$ for $i \in [4]$.
	
	\item To enforce minimum degree at least 3, we use sequential counters~\cite{Sinz05}.
	
	\item The clauses $ \bigvee_{v_2, v_3 \in [n] \setminus \{v_1\}, v_2 < v_3 } t_{v_1, v_2, v_3}$ for $v_1 \in [n]$ ensure that each vertex lies on a triangle.
\end{enumerate}
Additionally, we have definitional clauses to ensure that
$t_{v_1,v_2,v_3} \leftrightarrow (e_{v_1,v_2} \land e_{v_2, v_3} \land
e_{v_1, v_3} )$ holds.
\cv{\vspace{2pt}}

Whenever SMS returns a fully defined graph $G$ respecting the above constraints, we check if it is 010-colorable.
We use a simple backtracking algorithm for constructing 010-colorings. %
Given a 010-coloring~$c$ of a graph $G$, the clause \[ \bigvee_{c(v)=c(u) = 0} e_{v,u}  \lor \bigvee_{c(v_1)=c(v_2)=c(v_3) = 1 } t_{v_1,v_2, v_3} \] blocks all graphs colored by $c$. 

Our 010-coloring algorithm always attempts to color the vertices in a fixed order, but in case both colors are available for the next vertex, it uses a heuristic to determine the order in which the two colors should be tried.
This heuristics has access to edge frequencies among previously seen graphs, and prefers the color whose blocking clause will be less likely satisfied based on the frequency analysis.
The intuition behind this style of heuristic is this.
Edge distribution in canonical graphs is highly unbalanced; therefore different possible learnable clauses have a different likelihood of being falsified and leading to further conflicts, depending on which edge variables they contain.
We attempt to find colorings that will give rise to clauses where many literals will be falsified because their edges hardly ever occur in canonical graphs; we use the edge frequency statistics to guide the heuristic.

Taken to the extreme, we could, instead of computing a single heuristic 010-coloring, explicitly optimize in the space of all 010-colorings.
Our heuristic results are already good enough, and this is a bit out of the scope of this paper, but we believe interesting research questions may lie down this path.

\subsection{Parallelization} \label{sec:cubeAndConquer}

With our approach we are in fact able to verify KS non-existence for all $n \leq 22$ in time that is still practical for single-core sequential execution (see Table~\ref{table:candidates}).
In order to solve $n=23$, however, we will need to parallelize.

The most successful approach to parallelization in the context of combinatorial search problems is called \emph{cube-and-conquer}~\cite{HeuleKullmannBiere18}, where the original problem is split by applying mutually exclusive assumptions (\emph{cubes}) whose disjunction is implied by the original formula.
The main challenge with cube-and-conquer is finding cubes so that the subproblems are evenly hard.
In the original \emph{cube-and-conquer} paper, \citeauthor{HeuleKullmannBiere18}~use a look-ahead solver for generating the cubes. The problem with this approach in our case is that we would prefer cubes that take into account graph canonicity, which significantly skews the hardness and distribution of solutions in the search space, and which is not directly represented in the encoding itself. In order to construct representative cubes, we would effectively have to re-implement SMS in a look-ahead solver; we instead opted for the simpler option of using the same SMS solver for the cubing process as for the subsequent solving.

We proceed as follows for generating the cubes. We start the solver and whenever the number of assigned edge variables exceeds a pre-defined threshold, we add the partial assignment on the edge variables to the list of cubes, and exclude the current partial assignment on the edge variables from the solver. We continue until the solver concludes unsatisfiability. To further improve this approach, we first let the solver run for a specific amount of time (\emph{prerun}) before generating the cubes. The idea behind this is to let the solver learn some symmetry and coloring clauses first in order to get a more representative picture of the search space.  %

Note that with this technique the generated cubes are not necessarily mutually exclusive, i.e., one model can agree with several different cubes on the assigned variables. As we will see in our experiments, we can easily filter them.

Another important point is that the original encoding $F_n$ alone does \emph{not} imply the disjunction of our cubes, as would be the case in pure cube-and-conquer.
The reason is that we learn symmetry-breaking and coloring clauses along the way, and these further restrict the search space so that when the formula becomes unsatisfiable it is not just due to the cubes but also due to these additional clauses.
Instead, $F_n$ together with the learned symmetry-breaking and coloring clauses implies the disjunction of the cubes; we shall say $F_n$ entails the cubes \emph{with advice}.

\subsection{Embeddability Check}
The next step in the process of finding KS graphs is to check the candidate, non-010-colorable graphs for embeddability.
This check amounts to solving a first-order formula over multivariate polynomial equations over the real numbers, a problem that can be solved for example by the REDUCE algebra system~\cite{Dolzmann097} or the SMT solver Z3~\cite{Moura08}.
Because these equations can be hard, following previous work~\cite{LiBrightGanesh22}, we first check
whether the graph contains some known unembeddable
subgraph\footnote{\url{https://kochen-specker.info/smallGraphs/}
  provides a list of all unembeddable subgraphs with minimum degree 3
  up to order 14.}, and only if it does not, we resort to a full-blown Z3 embeddability check.
  We use the Glasgow Subgraph Solver~\cite{McCreeshP020} to test subgraph containment, being able to test all our candidates in less than an hour of CPU time in total.
  After this process, we are left with only two candidates which do not contain any known unembeddable subgraph of order $\leq 14$, and for those we have to fire up Z3.
  We next describe the encoding of embeddability, which also requires some non-trivial tricks.

Given a graph $G$ we want to decide whether we can find for each vertex three-dimensional real vectors so that adjacent vertices map to orthogonal vectors, and no two vectors are collinear.
We could simply write down these constraints in the language of Z3, but that would require 3 real-valued variables for the coordinates of each vertex, and seemed too difficult to solve when we tried it.
It turns out we can save a good chunk of the variables by substituting cross products for some of the vertices, as follows.

Since we are only interested in angles (orthogonality) and not magnitudes, we are effectively trying to map the vertices (injectively) to `directions' (1-dimensional subspaces) in $\mathbb{R}^3$.
Imagine we guess the directions for individual vertices one by one.
Once two vertices have their directions determined, any shared neighbor has only a single possible direction left, the direction given by the cross product.
It so happens that guessing the coordinates of only very few vertices often uniquely determines all other coordinates via cross products.
We formalize this phenomenon in the notion of a \emph{cross-product cover}.

We say $(S, w)$ is a cross-product cover of a graph $G$ if $S \subseteq V(G)$, $w$ maps each vertex $v \in V(G) \setminus S$ to a pair of its neighbors $w_v = \{w_v^1, w_v^2\}$, $vw_v^1, vw_v^2 \in E(G)$, so that the transitive closure of $\SB (u, v) \SM v \in V(G) \setminus S,\, u \in w_v \SE$ is anti-symmetric.
In other words, given the coordinates of vectors corresponding to the vertices in $S$, the vectors corresponding to the remaining vertices are uniquely determined by taking successive cross products, and there are no circular references. %
We call the vertices in $S$ and their vectors \emph{free}, and the others \emph{bound}.
Figure~\ref{fig:cross-product-cover} gives an example of a cross-product cover.

We write $p_v$ for the vector assigned to a vertex~$v$. Without loss of generality, we can assume that for one edge $v_1v_2 \in E(G)$ the corresponding coordinates are $p_{v_1} = (1,0,0)$, and $p_{v_2} = (0,1,0)$, provided that $v_1$ and $v_2$ are free vertices.

\begin{observation} \label{obs:embeddabilityEncoding}
	Let $G$ be a graph, $(S,w)$ a cross-product cover, $v_1v_2 \in E(G)$ an edge between free vertices. Then $G$ is embeddable iff there is a function $p: V(G) \rightarrow \mathbb{R}^3$ satisfying:%

\begin{enumerate}
	\item $p_{v_1} = (1,0,0)$, $p_{v_2} = (0,1,0)$; %
	\item $p_{v} \times p_u \not = 0$ for all $u,v \in V(G), u \not = v$: vectors are not collinear;
	\item $p_v = p_{w^1_v} \times p_{w^2_v}$ for all bound $v$, where $w_v = \{w_v^1, w_v^2\}$;
	\item $p_v \cdot p_u = 0$ for all $vu \in E(G)$: the inner product must be zero if $p_v$ and $p_u$ are orthogonal. 
\end{enumerate}
\end{observation}

\colorlet{triangle}{ijcaired}
\colorlet{starting}{azure}
\colorlet{arc}{azure}
\colorlet{edge}{gray}
\tikzstyle{Svert}=[fill=black,circle,scale=0.8]
\tikzstyle{Xvert}=[draw=black,circle,scale=0.8]
\setlength{\intextsep}{10pt}
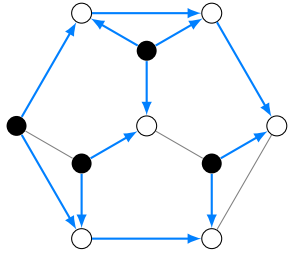
\begin{figure}[tb]
	
	\centering
	\begin{tikzpicture}[scale=0.5]
		\node[Svert] (T0) at (-1.73, -1) {} ;
		\node[Svert] (T1) at (-3.46,  0) {} ;
		\node[Xvert] (T2) at (-1.73, -3) {} ;

		\node[Svert] (S0) at (0, 2) {} ;
		\node[Svert] (S1) at (1.73,  -1) {} ;

		\node[Xvert] (C) at (0,  0) {} ;

		\node[Xvert] (O1) at (-1.73,  3) {} ;
		\node[Xvert] (O2) at ( 1.73,  3) {} ;

		\node[Xvert] (O3) at ( 3.46,  0) {} ;
		\node[Xvert] (O4) at ( 1.73,  -3) {} ;

		\draw[edge] (T0) edge (T1) ;
		\draw[thick,arc] (T0)  edge[-latex] (T2) ;
		\draw[thick,arc] (T1)  edge[-latex] (T2) ;
		\draw[edge] (S1) edge (C) ;
		\draw[edge] (O3) edge (O4) ;

		\draw[thick,arc] (T0) edge[-latex] (C) ;
		\draw[thick,arc] (S0) edge[-latex] (C) ;

		\draw[thick,arc] (T2) edge[-latex] (O4) ;
		\draw[thick,arc] (S1) edge[-latex] (O4) ;

		\draw[thick,arc] (T1) edge[-latex] (O1) ;
		\draw[thick,arc] (S0) edge[-latex] (O1) ;

		\draw[thick,arc] (O1) edge[-latex] (O2) ;
		\draw[thick,arc] (S0) edge[-latex] (O2) ;

		\draw[thick,arc] (O2) edge[-latex] (O3) ;
		\draw[thick,arc] (S1) edge[-latex] (O3) ;
	\end{tikzpicture}
	\caption{
		An illustration of a cross-product cover of a graph with ten vertices, one proved unembeddable by~\protect\citet{UijlenWesterbaan16}.
		Free vertices are in black, arrows indicate the cross-product relation.
		The remaining edges are drawn as simple lines.
	}
	\label{fig:cross-product-cover}
\end{figure}

Given a graph $G$, for any cross-product cover $(S,w)$ and an edge $v_1v_2 \in E(G)$ between free vertices, Observation~\ref{obs:embeddabilityEncoding} yields a set of constraints in polynomial real arithmetic that can be solved using Z3. Note that with the cross-product formulation we need variables only for free vectors (two of which are fixed), a total of only $3(|S|-2)$ real-valued variables. %
For most graphs we were able to find a cross-product cover with $|S|\leq 4$.
When we were unable to solve one set of constraints in 10 seconds we tried another cross-product cover.
We also tried normalizing the free vectors, but surprisingly this only made the problem harder.

\subsection{Comparison to SAT+CAS}

We have now laid out our process of search for KS systems.
Before we discuss experimental results and our new lower bound on the size of a KS system, let us compare SMS+CCL to the SAT+CAS approach recently used for the same problem~\cite{LiBrightGanesh22}.%
\footnote{The implementation of the latter is not available online.}

Both approaches have in common that they are based on SAT and are able to enumerate all graphs up to isomorphism described by an input encoding.
The first difference is in the canonical form: both produce lexicographically minimal graphs, but SMS produces lexicographically minimal graphs defined by concatenating the rows of the adjacency matrix whilst the SAT+CAS approach concatenates the above-diagonal entries of the columns. An important property of the latter is that the subgraph given by the first~$k$ vertices of a canonical graph is also canonical.

SMS uses a partition-based minimality check based on the partially defined graph which allows to construct symmetry breaking clauses even if only a few edges are assigned. In the SAT+CAS approach, one checks if the induced subgraph given by the first $k \leq n$ vertices is canonical, whenever all edge-variables between the first $k$ vertices are assigned.
A comparison between the amount of time spent in the solver and for the minimality check would be interesting.

The two approaches also differ in the way of ensuring that the resulting graphs are non-010-colorable. We add coloring clauses dynamically with CCL, while \citeauthor{LiBrightGanesh22}~added all coloring clauses for 010-colorings where at most $\lceil \frac{n}{2}\rceil$ vertices have color 1 upfront, leading to an exponential number of clauses.

The third difference is that \citeauthor{LiBrightGanesh22}~did not check for canonical form whilst cubing; they only used incomplete static symmetry-breaking constraints. This has the advantage that a different solver can be used, and they used the march-cu solver~\cite{HeuleKullmannBiere18}, but it foregoes the extra power of entailment with advice~(see Subsection~\ref{sec:cubeAndConquer}) and so might produce more cubes.
As we will see in our experiments, we enforce larger cubes in comparison to SAT+CAS; but since~\citeauthor{LiBrightGanesh22}~do not report the number of cubes, we cannot tell whether at the same time we produce more cubes. We suspect that SMS should be able to assign more variables thanks to the minimality check based on partially defined graphs.

\subsection{Computations}
\label{sec:KS-results}

In this section, we will describe our experimental setup. 
In the original SMS paper~\cite{KirchwegerSzeider21}, Clingo was used as solver. In our implementation, the user can choose between Cadical with the IPASIR-UP interface~\cite{FazekasNPKSB23} (a state of the art incremental CDCL SAT-solver with inprocessing) and Clingo (an ASP solver containing a CDCL SAT solver). For our experiments we decided to use Cadical.
The SAT encodings are created by a Python script and the
 embeddability check is also implemented in Python using Z3's Python
 interface. All our code is available on GitHub\footnote{\url{https://github.com/markirch/sat-modulo-symmetries}}.

We ran our experiments on a cluster with different processors\footnote{
Intel Xeon \{E5540, E5649,  E5-2630 v2,  E5-2640 v4\}@ at most 2.60 GHz, AMD EPYC 7402@2.80GHz}  under
Ubuntu 18.04. 
We use version 4.11.2
of Z3 and all tests are executed with a single thread.

We start with an experiment (Table~\ref{table:ks+ccl}) showing the impact of CCL for computing KS candidates by computing the number of graphs satisfying the existential part $F_n$ and the number of 010-colorings needed to filter out all 010-colorable graphs. Similarly to Table~\ref{table:CCL-vs-others-size}, we can see that the number of learned colorings is significantly lower than the total number of graphs satisfying the existential part, illustrating that CCL is a powerful technique for computing KS candidates.

\begin{table}[tb]
	\centering
	\begin{tabular}{@{}crr@{}}
		\toprule
		$n$ & \#existential  & \#colorings \\ \midrule
		13 & 34 & 3\\
		14 & 216 & 10 \\
		15 & 2352 & 32 \\
		16 & 27394 & 91 \\
		17 & 373646 & 267 \\
		18 & 6050114 & 832 \\
		\bottomrule
	\end{tabular}
	\caption{
		Comparison between the number of graphs satisfying the existential part $F_n$ of the encoding of KS candidates and the number of colorings computed to color all 010-colorable graphs.
		For $n=17$ there is one non-010-colorable graph among the 373646, all others counted in this table are 010-colorable.
	}
	\label{table:ks+ccl}
\end{table}

\begin{table}
	\centering
	\begin{tabular}{@{}crr@{~}l@{}}
		\toprule
		$n$ & \#KS candidates & \multicolumn{2}{c}{times} \\ \midrule
		17 & 1 & 12.93 sec & (1.2 min) \\
		18 & 0 & 71.67 sec & (7.8 min) \\
		19 & 8 & 8.43 min & (2.46 hrs)\\
		20 & 147 & 1.68 hrs & (39.71 hrs) \\
		21 & 2497 & 26.24 hrs & (42.56 days)\\
		22 & 88282 & 26.17 days & (5.26 years)\\
		\textbf{23} & \textbf{3747950} & \textbf{1.36 years} & \textbf{(not computed)}\\
		\bottomrule
	\end{tabular}
	\caption{Number of KS candidates. Time gives the total CPU time for solving this instance (including cubing for $n \in \{22,23\}$) and in brackets the times from \protect\citet{LiBrightGanesh22} ran on Intel E5-2683 v4@2.1GHz CPUs are given. The times for Clingo+SMS for $17\leq n\leq 20$ are 8.73, 70.47, 684,  and 3904 seconds respectively.} \label{table:candidates}
\end{table}

Table~\ref{table:candidates} summarizes the computation times for finding all KS candidates for a given order $n$ and gives the computation times provided from \citet{LiBrightGanesh22} for comparison. The number of KS candidates up to 22 vertices coincides with our computations.
For $n \in \{22,23\}$ we applied cube-and-conquer for parallelization, some details including prerun time (see Section~\ref{sec:cubeAndConquer}), the number of assigned edge variables, number of cubes, average solving time and longest time for a cube are given in Table~\ref{table:cubes}. Note that there is a large difference between the average time and the longest time for a cube. In the future, we want to investigate methods for generating more balanced cubes.

As already mentioned in Section~\ref{sec:cubeAndConquer} the KS candidates can have duplicates. This is indeed the case, for example for $n=23$ we got a total of 3752684 graphs, of which 4734 were duplicates, resulting in 3747950 unique graphs. %
Only two of them do not contain an unembeddable subgraph of size $\leq 14$ (we call them \emph{odd}).
We use Z3 to verify that these two are also unembeddable.
We thus prove Theorem~\ref{thm:KS24}.

\begin{theorem}
	\label{thm:KS24}
	 Every KS-graph has at least $24$ vertices.
\end{theorem}

\begin{table}
	\centering
	\begin{tabular}{@{}crrlcc@{}}
		\toprule
		$n$ & \#cubes & prerun & \#var. & $\overline{t}$ & $\max(t)$ \\ \midrule
		22 & 18659 & 2 days & 120 & 112 sec & 8946 sec \\
		23 & 313665 & 2 days & 140 & 137 sec & 63812 sec \\
		\bottomrule
	\end{tabular}
	\caption{Some details of constructing and solving the cubes for $n \in \{22,23\}$. Column ``\#cubes'' gives the number of cubes, ``prerun'' the prerun time before starting cubing, ``\#var.'' the lower-bound on literals for each cube, ``$\overline{t} $'' the average time for solving a cube, and ``$\max(t) $'' the time of the hardest cube to solve.} \label{table:cubes}
\end{table}

\subsection{Proofs for Computations}

Proving the absence of KS graphs of order $\leq 23$ involves several
computations. We discuss next how one can produce machine-verifiable
proofs for most of these steps.  For the results generated by SMS, we
can produce DRAT proofs~\cite{WetzlerHeuleHunt14} in a similar way as
described by \citet{KirchwegerScheucherSzeider22}, assuming
that the symmetry-breaking and coloring clauses are part of the
original encoding. We can check soundness of symmetry-breaking and
coloring clauses separately. To simplify this validation, we can store
permutations and colorings associated with these clauses.

As discussed in Subsection~\ref{sec:cubeAndConquer}, the disjunction of the cubes must be entailed with advice by the formula $F_n$ to ensure all models are accounted for. %
While constructing the cubes, we exclude each partial assignment corresponding to a constructed cube from the search space, and thus the cubing process ends with the formula \[\Omega = F_n \land (\lnot c_1 \land \ldots \land \lnot c_k) \land \Sigma \land \Gamma,\] where $\Sigma$ and $\Gamma$ are the learned symmetry-breaking and coloring clauses. Hence, a DRAT proof of unsatisfiability of $\Omega$ shows the completeness of the cubing process. Note that we may find some KS candidates already during the prerun before cubing; this is not a problem, a fully defined graph is just another cube.

Solving the formula under the assumption of a cube $c_i$ is equivalent to solving $F_n \land c_i$. During search we might find some KS candidates represented by the cubes $g_1, \ldots, g_t$ and exclude them from the search space, as well as learn symmetry clauses $\Sigma_i$ and coloring clauses $\Gamma_i$. Consequently, a DRAT proof of \[F_n \land c_i \land \Sigma_i \land \Gamma_i \land \lnot g_1 \land \ldots \land \lnot g_t \] shows that there are no more KS candidates for the assumption $c_i$.

We calculated an estimate of the proof size for $n = 23$ by
producing proofs for $5\%$ of the cubes. Assuming them to be
representative, the proof would have around $5$ terabytes.

When we prove a graph unembeddable because it contains a small known unembeddable subgraph, we can simply take the embedding given by the subgraph solver as a certificate.

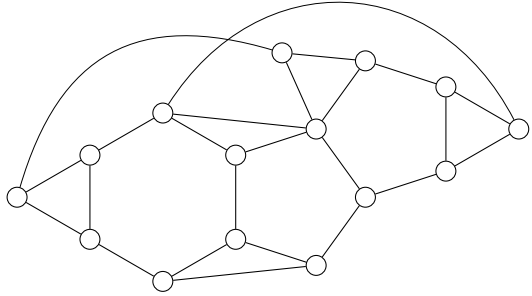
\begin{figure}
	\begin{center}
		\begin{tikzpicture}[scale=0.56]
		\node [Xvert] (0) at  (-4.46, 0)      {}; %
		\node [Xvert] (4) at  (+0.73, +1.00)  {}; %
		\node [Xvert] (3) at  (-1.00, +2.00)  {}; %
		\node [Xvert] (2) at  (-2.73, +1.00)  {}; %
		\node [Xvert] (1) at  (-2.73, -1.00)  {}; %
		\node [Xvert] (6) at  (-1.00, -2.00)  {}; %
		\node [Xvert] (5) at  (+0.73, -1.00)  {}; %
		\node [Xvert] (7) at  (+2.64, +1.62)  {}; %
		\node [Xvert] (8) at  (+3.81, +0.00)  {}; %
		\node [Xvert] (9) at  (+2.64, -1.62)  {}; %
		\node [Xvert] (10) at (+3.81, +3.24)  {}; %
		\node [Xvert] (11) at (+5.72, +2.62)  {}; %
		\node [Xvert] (12) at (+5.72, +0.62)  {}; %
		\node [Xvert] (13) at (+1.83, +3.43)  {}; %
		\node [Xvert] (14) at (+7.45, +1.62)  {}; %

		\draw (3) to (7);
		\draw (8) to (7);
		\draw (4) to (7);
		\draw (3) to (4);
		\draw (6) to (1);
		\draw (1) to (2);
		\draw (2) to (3);
		\draw (6) to (5);
		\draw (5) to (4);
		\draw (1) to (0);
		\draw (0) to (2);
		\draw (5) to (9);
		\draw (9) to (8);
		\draw (13) to (7);
		\draw (13) to (10);
		\draw (10) to (7);
		\draw (10) to (11);
		\draw (11) to (12);
		\draw (12) to (8);
		\draw (12) to (14);
		\draw (14) to (11);
		\draw [bend left=45,  looseness=1.25, line width=0.3] (0) to (13);
		\draw [bend left=300, looseness=1.25, line width=0.3] (14) to (3);
		\draw (6) to (9);
	\end{tikzpicture}
	\end{center}
	\caption{
		The shared 15-vertex subgraph of the two \emph{odd} graphs from Subsection~\ref{sec:KS-results}.
		Its graph6 string is \texttt{NGw@?i??GHgA@aCtQC?}.
	}
	\label{fig:15-subgraph}
\end{figure}

The final case is the full embeddability check when no known unembeddable subgraph is found. Verifying that a cross-product cover $(S, w)$ of $G$ is correct is straightforward, but unfortunately Z3 does not produce checkable proofs for nonlinear real arithmetic at the moment. %
We tested our implementation by deciding the embeddability of all square-free graphs with degree at least 2 up to order 12. The results coincide with previous work. Interestingly, we found two graphs claimed minimally unembeddable by previous works, from which an edge can be removed while preserving unembeddability. The graphs are still minimally unembeddable with respect to vertex removal, and also among graphs with minimum degree $3$.
Further, we extracted a common unembeddable subgraph of order $15$ from the two odd graphs by greedily removing as many vertices as possible while preserving unembeddability. 
This graph is shown in Figure~\ref{fig:15-subgraph}.

\section{Conclusion}

This work extended the SAT modulo Symmetries (SMS) framework with the co-certificate learning (CCL) technique for alternating search problems.
We showed that SMS+CCL can drastically reduce the search space compared to alternative methods, and demonstrated its potential by applying it to the prominent problem arising from quantum mechanics of determining the smallest Kochen-Specker (KS) graphs.
With our method we improved the best known lower bound and proved that KS graphs have at least 24 vertices.

A natural question is whether we can settle the case for $n=24$  with our current methods. We generated cubes for $n=24$ and tested 10\% of them; we expect a solving time of roughly 125 CPU years assuming that the solving time of the cubes is representative.
We plan to improve our current approach by excluding unembeddable partially defined graphs during the search. 

Another point for improvement is finding techniques for producing more balanced cubes, minimizing the solving time for the hardest cubes.

SMS+CLL is a general technique applicable to many problems. In recent work, we applied SMS+CLL to make progress on the Erd\H{o}s-Faber-Lovász Conjecture~\cite{KirchwegerPeitlSzeider23b} and there are many more problems which we plan to attack, for example, computing small non-2-colorable $n$-uniform hypergraphs~\cite{Ostergard14}, finding color-critical graphs with few edges~\cite{Jensen1995}, computing width-critical graphs for minor-closed width measures~\cite{Chlebikova02}.

\section*{Acknowledgments}
We acknowledge support from the Austrian Science Fund (FWF),
projects P32441 and J4361, and from the Vienna Science and Technology Fund (WWTF), project ICT19-065.

\av{\setlength{\bibsep}{6pt plus 0.5ex}}
\bibliographystyle{named}
\bibliography{ijcai23}

\end{document}